# The Effect of Structural Equation Modeling on Chatbot Usage: An Investigation of Dialogflow

Vinh T. Nguyen [a, *], Chuyen T. H. Nguyen [b]

[a] *Dept. of Information Technology, TNU – the University of Information and Communication Technology, Vietnam*
[b] *Dept. of Primary Education, Thai Nguyen University of Education, Vietnam*
*vinhnt@ictu.edu.vn, chuyennh@tnue.edu.vn*

| A R T I C L E   I N F O | A B S T R A C T |
|---|---|
|  | This study aims to understand users' perceptions of using the Dialogflow framework and verify the relationships among service awareness, task-technology fit, output quality, and TAM variables. Generalized Structured Component Analysis was employed to experiment with six hypotheses. Two hundred twenty-seven participants were recruited through the purposive non-random sampling technique. Google Forms was utilized as a medium to develop and distribute survey questionnaires to subjects of interest. The experimental results indicated that perceived ease of use and usefulness had a statistically significant and positive influence on behavioral intention. Awareness of service and output quality was considered reliable predictors of perceived usefulness. Also, perceived task-technology fit positively affected perceived ease of use. The model specification accounted for 50.04% of the total variation. The findings can be leveraged to reinforce TAM in future research in a comparative academic context to validate the hypothesis. Several practitioner recommendations and the study's limitations have been presented. |

* Corresponding author at:
  Dept. of Information Technology, TNU – the University of Information and Communication Technology, Vietnam
  Z115 Street, Quyet Thang commune, Thai Nguyen, 24000
  Vietnam
  E-mail address: vinhnt@ictu.edu.vn

ORCID ID:
- First Author: 0000-0002-1300-3943
- Second Author: 0000-0002-3846-8835

## 1. Introduction

We are living in the era of modern technologies where blockchain [1], the Internet of Things, artificial intelligence, and virtual/augmented reality [2][3][4] are pervasive in all aspects of life. Furthermore, the outbreak of the Covid-19 epidemic accelerated this development since individuals needed to use technology to communicate and work remotely [5]. As such, the number of software increased in such a short period in response to the needs of society. The advent of this software, on the one hand, delivers many benefits, but on the other hand, it also introduces many challenges and issues [6]. Some advantages include lower operational expenses, better use of time and resources, improved customer experience, individualized service, and the introduction of numerous new algorithms and programs. In addition, various issues and challenges remain that must be addressed in the future, such as infrastructure, service quality, massive data processing capacity, limited human resources, program/ algorithm usability, stability [7], or digital addiction [5][8].

The realm of education is not immune to technological changes along with the pandemic [9]. Existing software and tools, previously utilized for small and individual purposes, are now being used on a big scale with educators and students [6][7]. There is no denying the benefits such technologies provide, particularly in the context of the Covid-19 epidemic, as they play a significant role in connecting stakeholders, including teachers, learners, and policymakers. During this epidemic, numerous studies have revealed that many instructors and students are unaware of the availability of support resources [10]. They operate conventionally, from producing lectures, grading, and managing students to continuously updating the latest information [11]. Traditional methods of work will result in little productivity.

The artificial intelligence (AI) chatbot is a valuable tool for providing students with quick feedback, particularly in information management [12]. Microsoft Bot Framework, Dialogflow, IBM Watson, Pandorabots, and others are recently among the most widely used chatbot frameworks. Existing research on these frameworks has concentrated on technical factors such as how to build and apply them [12]. However, whether these frameworks were used in schooling throughout the epidemic is still being determined. To the best of our knowledge, research has yet to be undertaken to explore the adoption of AI chatbots, particularly Dialogflow (a natural language processing developed by Google), in administrative support tasks, leaving an unfilled gap in the literature.

The current study filled the research gap by posing the following research questions: (R1) what are the effects of service awareness, output quality, and perceived ease of use on perceived usefulness? (R2) to what extent can perceived effectiveness and ease of use predict behavioral intention to use Dialogflow? Moreover, (R3) to what time Task Technology Fit can expect perceived ease of use? The current study contributes to scarce research on AI chatbots in higher education by expanding the technology acceptance model (TAM) in the context of the overwhelming information burst during the Covid-19 epidemic, considering three external factors: awareness of service, output quality, and task technology fit.

## 2. Theoretical framework and hypothesis development

Numerous conceptual frameworks have been proposed and utilized in the literature for evaluating an IT application, including the Technology Acceptance Model (TAM), Information Systems Success Model (ISSM), and Uses and Gratifications Theory (UGT), so on. Each model was constructed based on several assumptions and factors. The current study adopted the parsimonious TAM model considering three additional factors: service awareness, output quality, and task technology fit.

Behavioral Intention (BI): The TAM was developed to probe the factors influencing customers' willingness to adopt the technology. Behavioral intention is the likelihood of a person performing some behavior [13]. The current study defines behavioral intention as the likelihood that teachers/learners will utilize Dialogflow for communication with each other. Three questions were proposed to measure behavioral intention, including

1. I plan on using Dialogflow in the next three months for communication,
2. I anticipate I will use Dialogflow in the next semester,
3. I plan to use Dialogflow to exchange information with students whenever necessary.

*Perceived Usefulness (PU):* Perceived usefulness is the belief of an individual that utilizing the system will help them acquire the job's objectives [13]. This research employs three questions to evaluate performance expectancy:

1. I would find Dialogflow useful for my communication.
2. I think using Dialogflow will improve the job's productivity.
3. I think utilizing Dialogflow will help me save administrative time.

The following hypothesis was proposed:
   **Hypothesis 1 (H1). Perceived usefulness has a positive effect on behavioral intention.**
*Perceived Ease of Use (PEU)*: The term ease of use is referred to the ease with which the system can be used [13]. It is an important predictor in the TAM model. This study adapted four questions to evaluate perceived ease of use, which are as follows:

1. It is easy for me to use Dialogflow.
2. It wouldn't need too much time for me to master Dialogflow.
3. Dialogflow interaction would be clear and concise.
4. Learning how to utilize Dialogflow would be straightforward for me.

The following hypotheses were posed:
   **Hypothesis 2 (H2). Perceived ease of use positively influences behavioral intention.**
   **Hypothesis 3 (H3). Perceived ease of use will have a positive effect on perceived usefulness**
*Task-Technology Fit (TTF)*: Task-Technology Fit assumes that the fitness of technology with the job will influence the current performance outcome [2]. Four questions were employed to assess Task-Technology Fit:

1. Dialogflow is adequate for creating and delivering the repetitive message,
2. Dialogflow is compatible with the task of exchanging information, and
3. Dialogflow is sufficient for automated communication:

The following hypothesis was proposed:
   **Hypothesis 4 (H4). Perceived Task-Technology Fit has a positive impact on perceived ease of use.**
*Awareness of Service*: This term implies users' awareness of a service's existence [14]. In the current study, the service refers to as Dialogflow [15]. Three questions were used to assess service awareness, which is as follows:

1. I heard Dialogflow from an advertisement,
2. I see my friends/colleagues use Dialogflow, and
3. I used Dialogflow before.

The following hypothesis was proposed:

**Hypothesis 5 (H5). Perceived awareness of service will affect perceived ease of use**.

*Output Quality*: Output quality refers to the extent to which an individual thinks the system performs the tasks well [16]. In the context of this study, output quality means users are satisfied with the feedback from Dialogflow. Three questions were proposed to assess output quality:

1. The quality of the output I received from Dialogflow is high.
2. I'm satisfied with the answers from Dialogflow.
3. Dialogflow provides me with appropriate responses.

The following hypothesis was proposed:

**Hypothesis 6 (H6). Output quality has a positive influence on perceived usefulness.**

## 3. Materials and Method

The subsections explain how hypothesis testing data is acquired, measured, and analyzed.

### 3.1. Data Collection

Non-probability purposive sampling was employed to acquire data for the study. Google Forms was a medium to create and distribute the online survey to subjects. Communication with participants for the survey was carried out via private messages and online social platforms (i.e., Facebook). The target population or participants of interest have previously used Dialogflow for communication. This study used snowball sampling to reach a diverse group of participants, beginning with the authors' networking channel. The research team asked peers and friends to spread the survey. Qualified individuals use Dialogflow at least once or, in other words, have experience with Dialogflow. The survey consists of two parts (a) 4 questions to acquire general information, (b) 19 Likert-type questions with a scale of (1 = Strongly Disagree, 2 = Disagree, 3 = Neutral, 4 = Agree, 5 = Strongly Agree) for different points of view using Dialogflow. The preceding part went into great depth on the 19 questions. This study revealed no identifiable personal information; hence no ethical approval was necessary.

### 3.2. Data Analysis

The current study employed Generalized Structured Component Analysis (GSCA) to confirm the influence of independent variables on the dependent variables proposed in the previous section [17]. GSCA is an alternative to PLS-SEM. Structural Equation Modeling (SEM) is a common approach used in social science to better get insights into the complex relationship among factors [18]. Covariance-based (CB-SEM) and component-based (PLS-SEM) are the two types of SEM. Both types require a large amount of data due to their standard distribution assumption [17][18]. However, GSCA does not involve normality issues. Therefore, it should impose fewer limitations on data distribution, produce distinctive component score estimates, and avoid wrong solutions in limited data (multivariate normality of observed variables is not required for parameter estimation) [17]. GSCA Pro [19] was used for hypothesis testing and additional data verification.

## 4. Result and Discussion

After data was obtained, a data cleaning process was performed where 41 invalid answers were removed due to consistent feedback (only select one option) and incomplete responses (missing values). Two hundred twenty-seven records were retained for the investigation (accounted for 82.46% of 268 responses). This study's sample size (227) exceeds the threshold recommended using sample estimator software (177).

### 4.1. Demographic characteristics

Data from Figure 1 shows that most respondents are males, accounting for 63.43%, while females only represent one-third of the sample size (36.57%). A possible explanation for this phenomenon is that men are more likely to experiment with and adopt new technology than women.

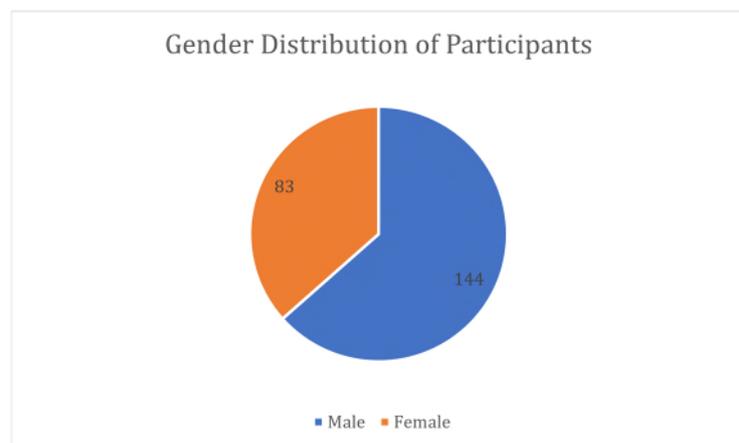

**Figure 1** Gender Distribution of Participants

In terms of age distribution, data from Figure 2 reveal that more than half of the respondents (156) are younger than 24 (68.72%), 23.78% are between the ages of 25 and 34, and 6.17% are between the ages of 35 and 44 and only three participants are over 44 years old.

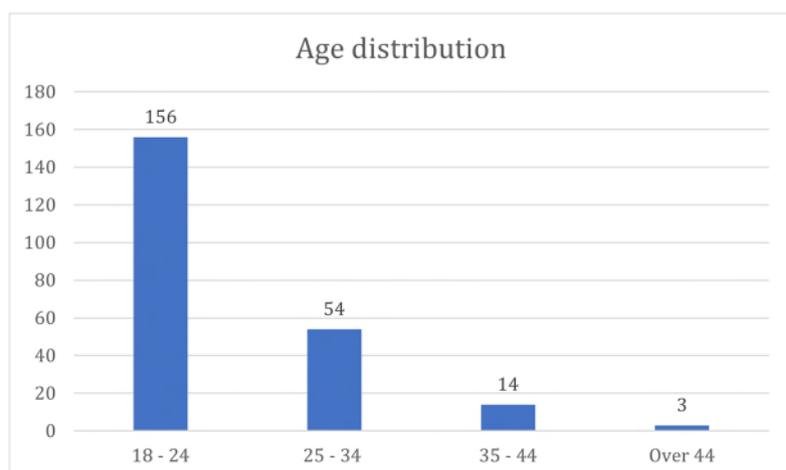

**Figure 2** Age Distribution of Participants

In terms of education level, 65.64% are undergraduate students, followed by being at the graduate level (30.40%), and the remaining respondents hold a Ph.D. degree (3.96%) (see Figure 3).

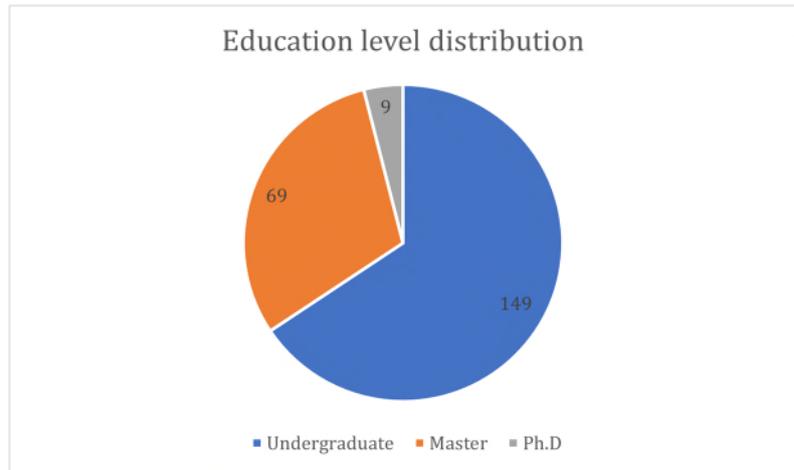

**Figure 3** Education Level Distribution

In terms of demographics, Figure 4 reports that almost two-thirds of participants live in rural areas (168/227), with the remainder residing in downtown (15.86%) and city areas (10.14%).

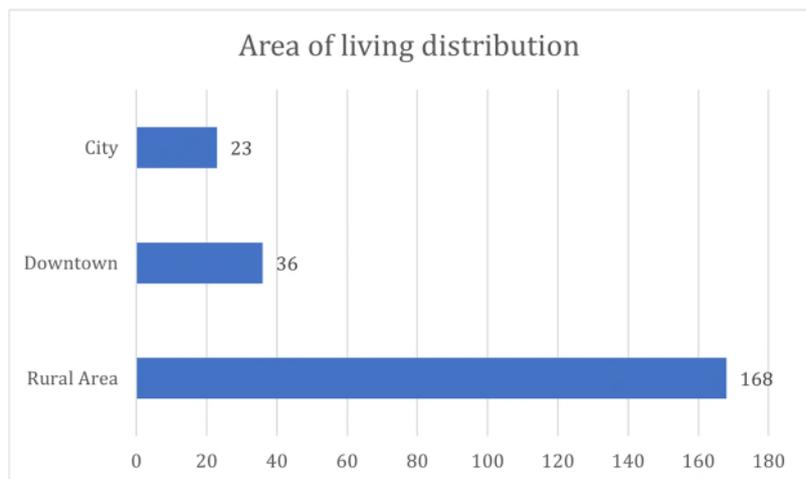

Figure 4 Area of Living Distribution

### 4.2. Quantitative Analysis

The mean and standard deviation for the construct items are shown in Table 1. The data highlighted that all of the extended TAM measures had means higher than the midpoint of 3 (min = 3.177, max = 3.682), with standard deviations between 0.528 and 0.848.

**Table 1** The Averages and Variations of the Variables (N =227)

| Construct | Item | Mean | SD |
|---|---|---|---|
| Awareness of Service (AW) | AWS1 | 3.55 | 0.81 |
| | AWS2 | 3.27 | 0.55 |
| | AWS3 | 3.17 | 0.56 |
| Output Quality (OQ) | OQ1 | 3.25 | 0.53 |
| | OQ2 | 3.31 | 0.58 |
| | OQ3 | 3.68 | 0.81 |
| Task Technology Fit (TTF) | TTF1 | 3.57 | 0.65 |
| | TTF2 | 3.66 | 0.81 |
| | TTF3 | 3.68 | 0.78 |
| Perceived Usefulness (PU) | PU1 | 3.40 | 0.76 |
| | PU2 | 3.54 | 0.78 |
| | PU3 | 3.46 | 0.81 |
| Perceived Ease of Use (PEU) | PEU1 | 3.34 | 0.81 |
| | PEU2 | 3.21 | 0.75 |
| | PEU3 | 3.44 | 0.85 |
| | PEU4 | 3.66 | 0.80 |
| Behavioral Intention (BI) | BI1 | 3.46 | 0.84 |
| | BI2 | 3.45 | 0.82 |
| | BI3 | 3.50 | 0.77 |

The internal consistency and convergent validity metrics for each construct are shown in Table 2. Each concept's internal consistency and reliability criteria were assessed using Dillon–Goldstein's rho [17]. The results ranged from 0.742 to 0.916. All exceeded the 0.7 thresholds for an acceptable reliability estimate [17]. The Average Variance Extracted (AVE) was also considered to examine whether it was convergent. All AVE values were more than 0.5, ranging from 0.526 to 0.784, showing that the convergent validity was reasonable.

**Table 2** Internal Consistency and Convergent Validity

| Construct | Item | Rho | AVE |
|---|---|---|---|
| Awareness of Service | 3 | 0.787 | 0.526 |
| Output Quality | 3 | 0.783 | 0.548 |
| Task Technology Fit | 3 | 0.916 | 0.784 |
| Perceived Usefulness | 3 | 0.742 | 0.589 |
| Perceived Ease of Use | 4 | 0.835 | 0.573 |
| Behavioral Intention | 3 | 0.887 | 0.723 |

Table 3 reported the loading estimates for the items, along with their standard errors (SEs) and 95% bootstrap percentile confidence intervals (CIs) with lower bound (LB) and upper bound (UB). The confidence intervals were calculated using 100 bootstrap samples. According to [17], a parameter estimate was considered statistically significant at the 0.05 alpha level if the 95 percent CI did not include a zero value. Evidence from data in Table 3 showed that there was no zero between the lower and upper bounds of any item, indicating reliable indicators of these items.

**Table 3** Estimate of Loadings

|      | Estimate | Std Error | 95% CI_LB | 95% CI_UB |
|------|----------|-----------|-----------|-----------|
| AS1  | 0.662    | 0.057     | 0.518     | 0.757     |
| AS2  | 0.665    | 0.077     | 0.511     | 0.762     |
| AS3  | 0.714    | 0.065     | 0.572     | 0.809     |
| OQ1  | 0.604    | 0.084     | 0.413     | 0.719     |
| OQ2  | 0.783    | 0.031     | 0.727     | 0.845     |
| OQ3  | 0.817    | 0.025     | 0.761     | 0.862     |
| TTF1 | 0.841    | 0.030     | 0.782     | 0.886     |
| TTF2 | 0.891    | 0.017     | 0.852     | 0.919     |
| TTF3 | 0.925    | 0.014     | 0.896     | 0.948     |
| PU1  | 0.737    | 0.056     | 0.583     | 0.836     |
| PU2  | 0.727    | 0.062     | 0.618     | 0.873     |
| PU3  | 0.643    | 0.119     | 0.231     | 0.764     |
| PEU1 | 0.853    | 0.022     | 0.802     | 0.889     |
| PEU2 | 0.779    | 0.042     | 0.681     | 0.847     |
| PEU3 | 0.875    | 0.016     | 0.834     | 0.899     |
| PEU4 | 0.437    | 0.076     | 0.279     | 0.571     |
| BI1  | 0.895    | 0.016     | 0.864     | 0.926     |
| BI2  | 0.856    | 0.019     | 0.822     | 0.895     |
| BI3  | 0.798    | 0.041     | 0.696     | 0.852     |

Table 4 reported model parameters' estimation from the GSCA approach, including FIT, Adjust FIT (AFIT), Goodness-of-Fit (GFI), and Standardized Root Mean Square Residual (SRMR). FIT indicates the total variance of all variables explained by the specified model. The value of FIT ranges between zero and one. The larger the FIT's value, the more variance is accounted for by the model. Here, FIT = 0.504 shows that the model specification explains 50.4% of the total variance. As reported in [17], There is no universal threshold for a FIT that indicates an acceptable fit. AFIT (0.499) is similar to FIT but considers model complexity. As another measure of overall model fit, GFI and SRMR show the closeness between sample covariance and covariance. GFI values around one and SRMR values near zero may be regarded as an indication of a good fit. The GFI value (0.926) was quite close to one, while the SRMR value (0.111) was rather significant and statistically different from zero.

**Table 4** Model FIT

| FIT   | AFIT  | GFI   | SRMR  |
|-------|-------|-------|-------|
| 0.504 | 0.499 | 0.926 | 0.111 |

Table 5 reported parameter estimation path coefficients in the specified model with standard deviation and 95% CIs. Generally, the explanations of the path estimated parameters are consistent with the interactions between the model's components. That is, perceived usefulness positively predicted behavioral intention (H1 = 0.412, SE = 0.115, 95% CI_LB = 0.088 and CI_UB = 0.578). Furthermore, perceived ease of use was verified to influence behavioral intention (H2 = 0.367, SE = 0.092, 95% CI_LB = 0.2, and CI_UB = 0.585). In addition, the effect of perceived ease of use on perceived usefulness was validated (H3 = 0.421, SE = 0.087, 95% CI_LB = 0.231, and CI_UB = 0.588). Moreover, perceived task-technology fit was confirmed to influence perceived ease of use (H4 = 0.636, SE = 0.063, 95% CI_LB = 0.495 and CI_UB = 0.744). In turn, perceived awareness of service had a statistically significant and positive impact on perceived usefulness (H5 = 0.195, SE = 0.063, 95% CI_LB = 0.03 and CI_UB = 0.303). Finally, output quality had a statistically significant and positive effect on perceived usefulness (H6 = 0.208, SE = 0.083, 95% CI_LB = 0.041, and CI_UB = 0.373).

**Table 5** Estimates of Path Coefficients.

|  | Estimates | Std Error | 95% CI_LB | 95% CI_UB |
|---|---|---|---|---|
| PU→BI (H1) | 0.412 | 0.115 | 0.088 | 0.578 |
| PEU→BI (H2) | 0.367 | 0.092 | 0.2 | 0.585 |
| PEU→PU (H3) | 0.421 | 0.087 | 0.231 | 0.588 |
| TTF→PEU (H4) | 0.636 | 0.063 | 0.495 | 0.744 |
| AW→PU (H5) | 0.195 | 0.063 | 0.03 | 0.303 |
| OQ→PU (H6) | 0.208 | 0.083 | 0.041 | 0.373 |

### 4.3. Discussion

#### 4.3.1. Theoretical implication

Perhaps, one of the most notable outcomes was the amount of variance explained by the extension of TAM with external factors (50.4%). The TAM model is a promising paradigm for examining this digital behavior. The results of the present research verified the plurality of the expected correlations between the factors in the proposed method. That is, perceived usefulness and ease of use had a statistically significant and positive influence on behavioral intention. This outcome verified the original TAM hypothesis and agreed with contemporary studies [20]. Thus, the data can supplement TAM to support the hypotheses in subsequent studies in a similar academic setting. The current findings also confirmed the theory that perceived ease of use predicts perceived usefulness. However, it supports Davis's work [21] and the present experiment [22]. In addition, Task-Technology Fit (TTF) is an external factor that has been incorporated into the model specification. The experimental results indicated that TTF positively influenced perceived ease of use. Its effects were aligned with similar studies [23][24], where TTF was considered a reliable predictor of perceived ease of use. What it means is that researchers might explore incorporating the TTF construct into their model to investigate digital technology behavior.

Regarding awareness of service, the experiment results showed a relationship between the awareness of service and perceived usefulness. Alternatively, in other words, perceived usefulness was positively influenced by service awareness. The current finding agrees with the hypothesis in another study [14]. Similarly, like TTF, service awareness might be a factor in understanding user behavior when employing technology. Finally, output quality had a statistically significant positive effect on perceived usefulness. This effect reinforces the findings in the study [14], in which the authors found that output quality is an essential predictor of perceived usefulness. As such, when researchers attempt to understand the direct influence on perceived usefulness or the indirect effect on behavioral intention, output quality may be included as a possible component in a theoretical study.

#### 4.3.2. Practical implication

The study's background is derived from the fact that, during the Covid-19 epidemic, instructors are required to perform much more administrative tasks in response to students' needs while working and learning remotely. Among many other chatbot software, Dialogflow is one of the most potential candidates to learn and apply in the current context. This is because Dialogflow is a product developed by Google and can be integrated with other Google services. Although this technology has been introduced for years, few users utilize its advantages. By understanding the causal relationship among factors that affect the intention to use such technology, this study sheds light on practitioners improving their product or customizing it to fit users' needs.

In terms of behavioral intention to use Dialogflow, the findings from the extended TAM model revealed that both perceived ease of use and perceived usefulness influenced it. Thus, while implementing or adapting Dialogflow in each scenario, software developers must verify that the program performs appropriately and is simple. In turn, ease to use is also the predictor of usefulness, meaning that practitioners should involve users during the development process (for example, considering the Agile model). Task-Technology Fit also contributes to its role in the model specification as it statistically influences perceived ease of use. As such, the software creator should identify users' tasks beforehand and utilize Dialogflow in such a way that it should fulfill the duties.

Furthermore, service awareness should not be disregarded because individuals may find the program valuable if they are aware of its presence. In this context, making Dialogflow available on many channels or incorporating it into educational courses might be promising. Finally, the output quality is a crucial component to consider since it directly affects the usability of an application or predicts the probability that users would utilize the investigated app. Therefore, practitioners should collaborate with users throughout the product life cycle to ensure that the output meets each individual's demands (and the Agile method is also a potential candidate in this case).

### 4.3.3. Limitation

Even though the results are founded on the efforts above, they will necessarily be limited by a variety of restrictions:

1. Non-probability sampling was employed in this research to ensure that participants had prior familiarity with chatbots. Despite widespread acceptance in the literature, this purposive sampling technique restricts generalization.
2. Because this study studied the adoption of Dialogflow within a short time, particularly in the context of the Covid-19 pandemic, the study's findings must be revisited after the outbreak to examine whether the same behavior still holds.
3. Other factors than those presented in the model specification are not considered.

Thus, further studies are called for investigation.

## 5. Conclusions

This study investigated the elements that affect people's intentions to use the Dialogflow framework by integrating task technology fit, service awareness, and output quality with the Technology Acceptance Model. Based on data from 227 participants, the study results confirmed the plurality of the hypothesized correlations between the parameters in the model specification. That is, perceived usefulness had a statistically significant and positive influence on behavioral intention. In turn, perceived ease of use had a statistically significant and positive impact on behavioral intention. In addition, perceived ease of use had a statistically significant positive effect on perceived usefulness.

Moreover, perceived task-technology fit had a statistically significant and positive influence on perceived ease of use. In turn, perceived service awareness had a statistically significant and positive impact on perceived usefulness. Finally, output quality had a statistically significant positive effect on perceived usefulness. The model specification explained 50.04% of the total amount of variance. The results can be used to supplement TAM to support the hypotheses in subsequent studies in a similar academic setting. Several recommendations for practitioners were discussed, along with the study's limitations.